\documentclass{article}

\usepackage{PRIMEarxiv}
\usepackage[utf8]{inputenc}
\usepackage[T1]{fontenc}
\usepackage{hyperref}
\usepackage{url}
\usepackage{booktabs}
\usepackage{amsfonts}
\usepackage{nicefrac}
\usepackage{microtype}
\usepackage{lipsum}
\usepackage{fancyhdr}
\usepackage{graphicx}
\graphicspath{{media/}}
\usepackage{authblk} 
\usepackage{tabularx}
\usepackage{enumitem}

\title{\large Exploring Physiological Responses in Virtual Reality-Based Interventions for Autism Spectrum Disorder: A Data-Driven Investigation

}

\author[(1,2)]{Gianpaolo Alvari}
\author[(1)]{Ersilia Vallefuoco}
\author[(1,2)]{Melanie Cristofolini}
\author[(3)]{Elio Salvadori}
\author[(3)]{Marco Dianti}
\author[(4)]{Alessia Moltani}
\author[(4)]{Davide Dal Castello}
\author[(1,2)]{Paola Venuti}
\author[(5)]{Cesare Furlanello}

\affil[(1)]{Department of Psychology and Cognitive Sciences, University of Trento, Rovereto, Italy}
\affil[(2)]{ODFLab, University of Trento, Rovereto, Italy}
\affil[(3)]{Meeva Srl, Trento, Italy}
\affil[(4)]{Comftech Srl, Monza, italy}
\affil[(5)]{HK3 Lab, Rovereto, Italy}

\begin{document}
\maketitle

\begin{abstract}
Virtual Reality (VR) has emerged as a promising tool for enhancing social skills and emotional well-being in individuals with Autism Spectrum Disorder (ASD). Through a technical exploration, this study employs a multiplayer serious gaming environment within VR, engaging 34 individuals diagnosed with ASD and employing high-precision biosensors for a comprehensive view of the participants' arousal and responses during the VR sessions. Participants were subjected to a series of 3 virtual scenarios designed in collaboration with stakeholders and clinical experts to promote socio-cognitive skills and emotional regulation in a controlled and structured virtual environment. We combined the framework with wearable non-invasive sensors for bio-signal acquisition, focusing on the collection of heart rate variability, and respiratory patterns to monitor participants behaviors. Further, behavioral assessments were conducted using observation and semi-structured interviews, with the data analyzed in conjunction with physiological measures to identify correlations and explore digital-intervention efficacy.
Preliminary analysis revealed significant correlations between physiological responses and behavioral outcomes, indicating the potential of physiological feedback to enhance VR-based interventions for ASD. The study demonstrated the feasibility of using real-time data to adapt virtual scenarios, suggesting a promising avenue to support personalized therapy. The integration of quantitative physiological feedback into digital platforms represents a forward step in the personalized intervention for ASD. By leveraging real-time data to adjust therapeutic content, this approach promises to enhance the efficacy and engagement of digital-based therapies.
\end{abstract}

% keywords can be removed

\section{Introduction}

Autism Spectrum Disorder (ASD), as delineated in the Diagnostic and Statistical Manual of Mental Disorders, Fifth Edition (DSM-5, \cite{american2013diagnostic}), manifests through distinct patterns of behavior and challenges in social communication and interaction. This includes impairments in social-emotional reciprocity, atypical social approaches, difficulties in initiating and maintaining conversations, and challenges in developing and maintaining relationships. These social deficits often deepen during the transition to adulthood, exacerbating issues related to daily living, interpersonal relationships, loneliness \cite{grace2022loneliness}, depression \cite{pascoe2023exploring}, and employability \cite{solomon2020autism}.\\
Emerging research has highlighted virtual reality (VR) as a promising avenue for enhancing specific skills in individuals with ASD, particularly adolescents \cite{glaser2022systematic,parsons2002potential}. VR technology facilitates the creation of simulated 3D environments, offering users an immersive experience that varies in intensity based on the technology employed. From non-immersive desktop environments to fully immersive virtual worlds \cite{jerald2015virtual,anthes2016state}, VR's potential to engage and motivate users is often augmented through the integration of serious game (SG) framework. Moreover, VR provides a secure, controlled setting for individuals with ASD to interact with reduced social anxiety, allowing them to explore and practice various social scenarios \cite{parsons2015learning,zhang2020assessing}.\\
Previous studies \cite{kim2024promoting,wade2017pilot,garzotto2018exploiting,simoes2020virtual,kotsopoulos2021vress} have integrated VR applications with physiological data to monitor the emotional state of individuals with ASD and to assess stressful or anxiety-provoking situations in virtual scenarios. Specifically, wearable non-invasive sensors were used to collect and record physiological parameters such as heart rate, skin conductance, and respiratory rate. Variations in these physiological responses correlate with the body's response to the release of stress hormones \cite{migovich2024stress}. For instance, as reported by \cite{cheng2020heart}, individuals with ASD exhibit significantly lower values of heart rate variability (HRV) reactivity in stressful social situations than typically developing peers. Variations in breathing frequency may be related to the challenges individuals with ASD face in regulating and stabilizing their physiological responses following exposure to stressful events \cite{thoen2023differences}.\\
Although extensive research has been carried out on VR-based training for individuals with ASD, the majority of studies have predominantly focused on the of training various social abilities  (e.g., job-related social skills \cite{burke2018using,smith2015brief}, social communication skills, \cite{li2023faceme,herrero2020immersive} or providing scenarios of dyadic interactions \cite{mesa2018effectiveness}). Few studies \cite{zhang2020assessing,zhao2018hand,ke2018virtual,babu2020multiplayer} have used collaborative immersive virtual environments that enable multiple users to interact and cooperate in real-time. In addition, to the best of our knowledge, no research has been found that designed VR applications for ASD group intervention, considering the therapist as a mediator and facilitator of the proposed activities. \\
This study illustrates a feasibility study aimed to investigate the effects of group VR gaming on social skills and emotional well-being in pediatric patients with ASD. The VR-based multiplayer SG was developed by Meeva and presented in \cite{gabrielli2023co}. The goal of SG is to enhance social skills and executive functions through several collaborative activities. In this study, we integrate the SG with a physiological-based monitoring system to further investigate interactions and responses of individuals with ASD. In particular, 34 participants with ASD were enrolled in the study. They engaged in VR gaming sessions in small groups (maximum of 4 people) for at least one session, with a therapist. Behavioral responses of participants were assessed by two independent raters using an observational form, while physiological data were collected to evaluate the emotional well-being of participants through the monitoring system. A comprehensive analysis was conducted to explore the relationship between behavioral and physiological variables, with the aim of assessing the feasibility of this approach and to contribute to a more comprehensive understanding of ASD within interactive VR environments.
The remaining sections of the paper are as follows: the Methods section outlines the characteristics of the participants, describes the VR platform developed for the study, outlines the procedure of the experimental sessions, presents the measured outcomes, and details the data analysis. The Results section presents the results of the analysis of the behavioral and physiological measures and explores the correlations between these measures. The Discussion section discusses the results in the context of the current literature.

\section{Methods}
\label{sec:headings}

\subsection{Participants}

This study involved 34 participants diagnosed with ASD (mean age = 11.7 years; 28 males and 6 females), recruited during the "Terapia in Vacanza" summer camp (Trento, Italy), an initiative organized by the University of Trento's Laboratory of Observation, Diagnosis, and Education (ODFLab), in partnership with the Albero Blu socio-health cooperative. Eligibility criteria included a formal ASD diagnosis according to DSM-5 criteria \cite{american2013diagnostic}, an intelligent quotient (IQ, WISC-IV, \cite{kaufman2006test}) score of 70 or above, no significant language impairments, and provision of consent to participate in the study. Exclusion criteria encompassed the absence of an ASD diagnosis per DSM-5, concurrent impairments, severe psychiatric conditions, and lack of consent to participate in the study. Participant demographics and characteristics are detailed in Table~\ref{tab:participant_characteristics}. Consent was obtained from all participants and their guardians. The Ethical Committee of Azienda Provinciale Servizi Sanitari (Trento, Italy) approved the study (protocol number: 4959, dated March 15, 2023).

\begin{table}[ht]
\centering
\caption{Characteristics of Participants by Age Group}
\label{tab:participant_characteristics}
\begin{tabular}{p{8cm}p{3cm}p{3cm}} % Adjust the width of the first column here
\toprule
& \textbf{Pre-Adolescents (8-12 years)} & \textbf{Adolescents (13-18 years)} \\
\midrule
\textbf{N} & 18 & 16 \\
\textbf{Age (months), mean (SD)} & 115.6 (23) & 179.8 (19) \\
\textbf{Males (\%)} & 84.5\% & 89.4\% \\
\textbf{ADOS (ADOS-2 comparison score), mean (SD)} & 6.4 (1) & 7.0 (1) \\
\textbf{ADOS Tot (ADOS-2 total raw score), mean (SD)} & 11.5 (2) & 13.4 (4) \\
\textbf{ADOS SA (ADOS-2 Social Abilities raw score), mean (SD)} & 8.6 (2) & 10.7 (3) \\
\textbf{IQ (WISC-IV Intelligence Quotient), mean (SD)} & 94.4 (21) & 100.6 (13) \\
\textbf{VCI (WISC-IV Verbal Comprehension Index), mean (SD)} & 95.3 (26) & 104.6 (15) \\
\textbf{PRI (WISC-IV Perceptual Reasoning Index), mean (SD)} & 109.3 (23) & 119.8 (15) \\
\textbf{WMI (WISC-IV Working Memory Index), mean (SD)} & 85.1 (14) & 87.8 (14) \\
\textbf{PSI (WISC-IV Processing Speed Index), mean (SD)} & 87.7 (20) & 79.8 (11) \\
\textbf{Duration of the session (minutes), mean (SD)} & 28.3 (5) & 29.2 (4) \\
\bottomrule
\end{tabular}
\end{table}

\subsection{VR-based Serious Game}

The SG is a collaborative multiplayer VR game and has been released as an application for Meta Quest 2. It was designed to provide an alternative rehabilitation scenario for therapists to carry out group interventions. In fact, the therapist is a player who supports other players where necessary and customizes the gaming experience, such as sensory inputs and difficulty level. The game requires the players to save the planet by completing several collaborative activities proposed in three game levels: 1) Coin Search Scenario (Coin): players have to search coins in a virtual park; 2) Station Scenario (Station): players have to buy a ticket with the collect coins to reach the planet where found the materials useful to save the planet; 3) Battle Scenario (Battle): players have to shoot rocks to find the material. The SG aims to support the achievement of specific therapeutic objectives:
\begin{itemize}
    \item Cognitive functions: this includes sustained attention, defined as the ability to maintain an instruction until the task is fully completed; selective attention, the ability to focus on one or more elements while ignoring distractors; single inhibition, the ability to refrain from a behavior in favor of another required behavior; and working memory, the capacity to retain a recently referred element.
    
    \item Socio-relational aspects: this covers collaboration, the ability to work as a team to complete a task; social openness towards others, the capacity to communicate or interact with another person; social response, the ability to respond to an-other's approach; and turn-taking, the ability to acknowledge an-other's presence and recognize and respect their role in the interaction.
\end{itemize}

The design and the complete description of the game is reported in \cite{gabrielli2023co}.

\subsection{Bio-signal Acquisition and Preprocessing}

\begin{figure}[ht]
    \centering
    \includegraphics[width=0.6\textwidth]{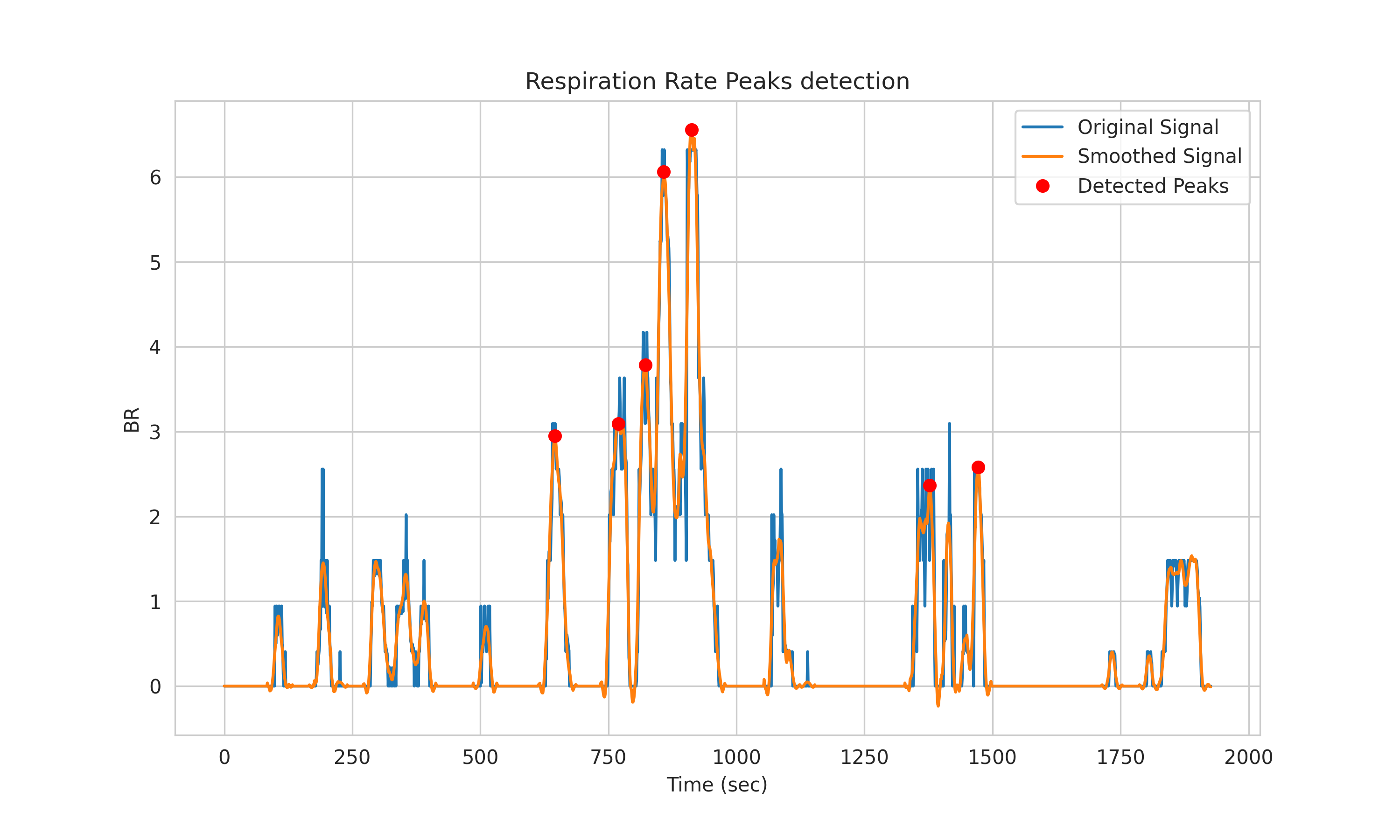} % Replace 'example-image' with the filename of your image
    \caption{Respiration Peaks Extraction (Artificial Session Example)}
    \label{fig:peaks}
\end{figure}

During the game sessions, each participant's level of activation was monitored using high-precision patented biosensors: Comftech HOWDY \cite{lavizzari2021heart,bizzego2021improving}. This device consists of a wearable textile band equipped with three electrodes (two abdominal, one dorsal) with a conductive surface in direct contact with the skin and a connection device. The sensor connects via Bluetooth to a specific Android mobile application, allowing for continuous signal acquisition at a sampling rate of 128 Hz. The device can detect various bio-signals, including the ECG signal, and respiratory signal. The application then transmits the data stream to a cloud server, which maintains each user's sessions and saves the signals in JSON files.\\
In the experimental phase, each participant was assigned a HOWDY sensor and a recording device. They were asked to wear the moistened bands a few minutes before the start of the activity to ensure signal acquisition throughout the session and allow the participants to acclimate. Therapists could monitor the acquisition status in real-time through the mobile application, ensuring that the signal was sufficiently stable. At the beginning of the activity, the first few minutes of relaxation allowed for the recording of the baseline, enabling the normalization of the activation level based on individual characteristics. Within the sample, four subjects refused to wear the bands during the intervention sessions, while the remaining 30 agreed.\\
For each monitored participant, three physiological signals were extracted: heart rate, intra-beat intervals, and respiratory rate. The heart rate (HR) and RR intervals (RR) are calculated from the ECG signal, while the breathing frequency (BF) is calculated from the respiratory signal \cite{shaffer2017overview}. The bio-signals were pre-processed in Python using libraries (SciPy, pyphysio, \cite{bizzego2019pyphysio}) designed for physiological signal processing, applying filtering, resampling, interpolation, and normalization. In the first step of the pipeline, the data were filtered through the application of a Low-Pass filter (LPf) and resampled at 1Hz. The signals were then interpolated using Random Forest regression to estimate missing values caused by errors or sensor disconnection, obtaining a continuous and uniform signal. During the interpolation step we excluded sessions with over 50\% of missing values (n=33). Finally, the values were normalized using winsorized normalization to reduce the effect of outliers by replacing them with the maximum and minimum non-outlier values. Specifically, for each session signal segments before the activity corresponding to the baseline phase (duration M=119 sec) were first extracted by manually coding from video-recordings (BORIS, \cite{friard2016boris}) of the interaction before the activity, when participants were seated and calm; outliers were then removed by cutting 5\% of the lowest and highest values and replacing them with the 5th and 95th percentile values, respectively. The trimmed segment was then used for z-score normalization.\\
Following the preprocessing phase, various variables related to the three types of signals were extracted. Specifically, (1) from the HR signal, variables in the time domain were extracted, including the mean and Standard Deviation (SD) of heart rate (HR SD, bpm), the Coefficient of Variation (CV), which is calculated from the ratio between these last two and represents a measure of relative variability, and Kurtosis, which describes the shape of the distribution of values relative to a normal distribution. Regarding (2) the RR signal, in the time domain were calculated: mean and SD of RR intervals, CV, and Kurtosis. Additionally, to obtain a more specific measurement of Heart Rate Variability (HRV), the following were extracted: SD of successive differences (SDSD), root mean square of successive differences (RMSSD), the percentage of pairs of adjacent NN intervals differing by more than 50 ms (pNN50), and SD of NN intervals (SDNN). In the frequency domain, the frequency spectrum was initially calculated using the Welch method employing 256-second windows with 50\% overlap, from which were calculated: Total Power (TP), Low-Frequency power (0.04-0.15 Hz, LF), High-Frequency power (0.15-0.4 Hz, HF), and the ratio of Low to High-Frequency power (LF/HF). Finally, (3) from the BF signal, the mean and SD of respiratory rate, CV, and Kurtosis were extracted. To further analyze the BF signal, we employed a custom algorithm combining a Savitzky-Golay filter for noise reduction without compromising signal trends, and a peak-finding function (SciPy) for feature extraction related to respiratory rate peaks. The filter  was applied with a specific 30-sec window and a polynomial order of 5, in order to optimally preserve the attributes of breathing cycles. Then we applied the peak-finding function on the smoothed signal using a height threshold of 2*SD and a prominence threshold of 0.5*SD (Figure \ref{fig:peaks}). By implementing this approach we were able to extract: rate (PRate), average prominence, and average widths of respiration peaks.

\subsection{Behavioral Assessment Tools}

To evaluate participants' experiences across the game sessions, ad hoc instruments were created, outlining the objectives investigated in each scenario, along with their tasks and functions. Observational form was developed to investigate behaviors (total features n=18) and experiences in relation to two macro-areas:
\begin{itemize}
    \item VR game experience  (5 features): VR adaptation difficulties, visuo-spatial difficulties, involvement in the activity, quality of relationship between peers, attention to instructions;
    \item Social interaction  (4 features): social openness towards peers and the therapist, social reciprocity with peers and the therapist.
\end{itemize}
To evaluate and measure the extent of social interactions, the therapist tracked the occurrence of social behaviors under 4 distinct conditions: Spontaneous, Verbally Suggested, Verbally Indicated, and Prompted. Spontaneous behavior occurred without any prompt or suggestion; Suggested refers to behaviors encouraged indirectly by the therapist; Indicated refers to behaviors explicitly requested by the therapist; Prompted includes the sum of behaviors both suggested and explicitly indicated. The frequencies of these behaviors were normalized based on the duration of the activity, yielding 4 rates for each of the 4 features studied, except for the social openings to the therapist (SO Therapist), which naturally could only be categorized as spontaneous (total social interaction features n=13). 
The observation form was completed by two experienced observers by watching videos of the game sessions. Cohen's kappa coefficient was calculated to assess the degree of agreement between the observers who completed the observation form, which indicated almost perfect agreement (\( \kappa > 0.80 \)) between the two observers. 
To assess VR game overall adaptation and difficulties (n=4 features), a 5-point Likert scale was also used for each feature, based on therapists' perceptions of participants' experience during VR game experience as follows: 0= easily; 1=fairly easily; 2 = neutral; 3= fairly difficulty; 4=difficulty. In addition, to assess the user VR experience, a brief structured interview \cite{gabrielli2023co} was administered to participants.

\subsection{Procedure: Intensive Therapy Week Experience}

During the intensive therapy week, each participant (N=34, mean age=11.7 years, SD=3) engaged with the SG in VR for at least 1 session (N=32, mean age=12.6 years, SD=4), at least 2 sessions (N=24, mean age=11.1 years, SD=3), and at least 3 sessions (N=20, mean age=10.9 years, SD=3) on different days. Each session, lasting approximately 30 minutes (mean duration=29 min, SD=4), was video recorded and coded using the observation form by two external observers.\\
Participants were divided into groups of 2 or 3 based on age and ASD severity level. The groups remained constant across gaming sessions and participated in multiplayer mode with a consistent therapist present to mediate the game interaction. Prior to the sessions, an agreement on instructions to be given to participants during the game was established. To ensure gradual exposure to the equipment and familiarization with the virtual environment, the first session included an explanation phase and a VR familiarization phase. In line with \cite{whyte2015designing}, each session gradually introduced a new scenario: during the first session, players only played the first scenario (Coin); in the second session, the scenarios Coin Search and Station; and in the third session, all scenarios, including Coin Search, Station, and Battle. Moreover, sessions for each group were conducted at the same time of day to minimize variability related to factors such as fatigue or differences in activities performed before the VR sessions.\\
Additionally, aspects related to well-being were evaluated through gradual exposure to the virtual environment, controlled and personalized by the therapist for different subjects, detected through qualitative observation by external observers present during the sessions and through a specifically designed interview administered at the end of the week

\subsection{Data Analysis}
The data analysis was conducted in several stages, incorporating both qualitative and quantitative approaches to comprehensively understand the impact of the VR intervention on participants with ASD and exploring the physiological dynamics within the sessions. Throughout all phases of analysis, outliers (values distributed beyond 3 standard deviations from the mean) were filtered and removed.

\subsubsection{Analysis of Behavioral Variables and Control Factors}
The relationship between behavioral, physiological and clinical features and control variables (age, sex) was examined. Spearman's correlation and point-biserial correlation tests were utilized for continuous or ordinal variables and dichotomous variables, respectively, following guidance by \cite{schober2018correlation} for interpreting correlation coefficients in clinical research.

\subsubsection{Game Session and Scenario Comparisons}
Dynamics of behavioral and physiological variables across three game sessions and scenarios were compared using a Friedman test for repeated measures, followed by Wilcoxon pairwise tests for post-hoc analysis, to understand the variability and consistency across different game interactions.

\subsubsection{Relationship Between Behavioral and Physiological Variables}
To explore the relationship between physiological and behavioral data, our analysis focused on the sample at the session level rather than the individual subject level, considering data collectively from all sessions (N= 64). We divided this step into 2 approaches:

\begin{enumerate}
    \item[\textbf{A.}] \textbf{Physiological-Behavioral Correlations}\\
    Generalized Linear Models (GLMs) were employed to explore the relationships between physiological-behavioral and physiological-clinical variables, accommodating non-normal distributions. Distribution characteristics were assessed through skewness, kurtosis with Fisher's adjustment, the Shapiro-Wilk test, and QQ-plots. This informed the selection of appropriate GLM types: Poisson or Gamma models for skewed behavioral variables. Models were evaluated using the Akaike Information Criterion (AIC) and analysis of residuals distribution, with variables standardized before fitting.
    
    \item[\textbf{B.}] \textbf{Unsupervised Analysis of Behavioral and Physiological Relationships}\\
    An unsupervised approach identified potential behavioral subgroups within the clinical cohort. 
    Behavioral features, alongside age—a variable chosen for its proven role as a relevant factor—and subject ID, were employed to preserve the individuality of the subjects in the analysis. Variance filtering removed behavioral variables with low variability (below the 50th percentile). After standardizing the remaining variables, we employed t-distributed Stochastic Neighbor Embedding using (t-SNE) with Barnes-Hut approximation for dimensionality reduction to 2 components. For unsupervised clustering we used KMeans to split the sample into n clusters, using Silhouette and Davies-Bouldin scores to find the optimal number. Ultimately, the validation of the clusters was performed by analyzing demographic (age, sex), clinical, and physiological variables, to highlight characteristics of each subgroup. The comparison between each pair of  cluster characteristics was conducted using the Mann-Whitney U test, ensuring a statistical evaluation of their differences.
\end{enumerate}

\section{Results}

Given the abundance of results, for clarity only statistically significant results (p<.05) are reported below.

\subsection{Behavioral Variables and Control Factors}

For sex, analyses were conducted using the point-biserial correlation test, revealing significant relationships across several features. Prompted social responses to the therapist (\textit{SR Therapist}, $r = .21$, $p = .018$) and to peers ($r = .18$, $p = .044$) both indicated significant sex differences, with females showing higher mean rates. Considering the clinical profile, the Perceptual Reasoning Index (PRI) was significantly associated with sex ($r = .26$, $p = .020$), where females had higher mean scores. The Working Memory Index (WMI) revealed a sex-related difference ($r = .41$, $p < .001$), with females outperforming males. No difference was found considering the physiological profile collected during the sessions. \\
Spearman's rho was employed to examine the relationship between age and various measures. A negative correlation was found between age and HR average ($\rho = -.21$, $p = .028$), indicating lower heart rates in older participants. Similar negative correlations with age were observed for RMSSD ($\rho = -.20$, $p = .036$) and SDSD ($\rho = -.20$, $p = .035$), suggesting reduced HRV in older individuals. HF power also showed a negative correlation with age ($\rho = -.22$, $p = .021$), while the LF/HF ratio in RR intervals exhibited a positive correlation ($\rho = .25$, $p = .007$), suggesting changes in autonomic nervous system balance with age. The rates of spontaneous social openings to peers (SO Peers) displayed a significant negative correlation with age ($\rho = -.29$, $p < .001$). Reciprocal social responses to the therapist (\textit{SR Therapist}) were negatively correlated with age ($\rho = -.19$, $p = .031$), whereas the overall quality of the relationship with peers (Relation PP) had a positive correlation ($\rho = .26$, $p = .003$), indicating nuanced shifts in social behavior and peer relationships over time. \\
These findings seem to delineate significant patterns of how sex and age influence various aspects of behavior and physiological responses, with sex differences noted in social initiatives and comprehension, and age-related shifts in physiological markers and social behaviors.

\subsection{Game Session and Scenario Comparisons}

\begin{figure}[ht]
    \centering
    \includegraphics[width=0.8\textwidth]{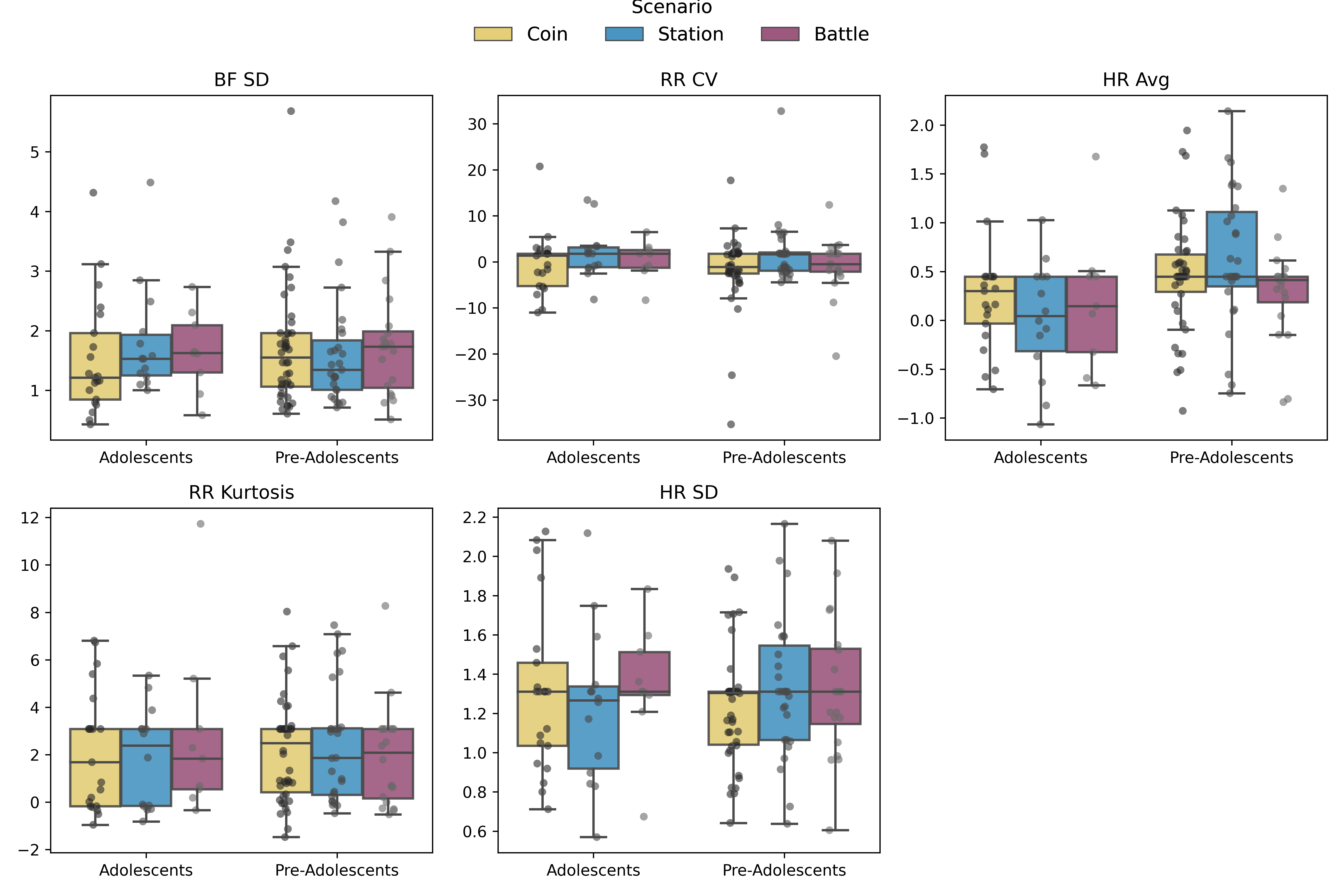} % Replace 'example-image' with the filename of your image
    \caption{Physiological features from significant comparison between scenarios (Coin, Station, Battle), divided by age sub-samples (Adolescents, Pre-Adolescents)}
    \label{fig:scenario}
\end{figure}

Given the significant association between age and various physiological and behavioral features, our analysis divided the sample into two age categories: adolescents (156 months and older) and pre-adolescents (younger than 156 months). This categorization allowed for nuanced comparisons within each age group across different VR game scenarios: Coin, Station, and Battle. We focus on the results from the Friedman test and subsequent post-hoc Wilcoxon tests, incorporating directionality indicated by the average values to elucidate the relationship between scenarios (Figure \ref{fig:scenario}). For clarity, the results are presented listing significant results for Adolescents sub-sample first, and then Pre-Adolescents.\\

\begin{itemize}
    \item \textbf{Adolescents:}
    \begin{itemize}
        \item Respiration rate SD displayed variability across scenarios ($\chi^2(2) = 7.79$, $p = .020$). The significant difference between Coin and Station scenarios ($Z = 48.0$, $p = .019$) and Coin and Battle ($Z = 59.0$, $p = .049552$) aligns with an increase in average values from Coin ($M=1.746$) to Station ($M=1.91$) and Battle ($M=1.89$).
        \item For RR CV, significant changes were identified ($\chi^2(2) = 7.24$, $p = .027$). A significant difference was observed between the Coin and Station scenarios ($Z = 51.0$, $p = .044$), with the average coefficient of variation increasing from the Coin scenario ($M = 0.71$) to the Station scenario ($M = 1.81$). This suggests a higher variability in heart rate during the Station scenario.

        \item Overall quality of the relationship between peers (Relation PP) varied significantly across VR game scenarios ($\chi^2(2) = 8.12$, $p = .017$). Significant differences were observed between the Coin and Station scenarios ($Z = 25.0$, $p = .007$) and between Coin and Battle scenarios ($Z = 18.0$, $p = .002898$). The average score for peer relationships decreased from the Coin scenario ($M = 2.76$) to the Station ($M = 2.26$) and further to the Battle scenario ($M = 2.12$).
        \item Social openings to therapists (SO Therapist) varied significantly across VR game scenarios for adolescents ($\chi^2(2) = 13.162$, $p = .00139$). Higher rate of openings was found in Station ($M=3.388$) respect to Coin ($M=2.113$, $Z = 23.0$, $p = .0013$) and Battle ($M=2.4459$, $Z = 20.0$, $p = .0412$).

    \end{itemize}

    \item \textbf{Pre-Adolescents:}
    \begin{itemize}
        \item Social openings to therapists (SO Therapist) varied significantly across VR game scenarios ($\chi^2(2) = 7.11564$, $p = .0285$). Higher rate of openings was found in Station ($M=2.97$) with respect to Battle ($M=2.17$, $Z = 68.0$, $p = .01099$).
        \item VR adaptation difficulties (Adaptation Diff) varied significantly across VR game scenarios ($\chi^2(2) = 7.1724$, $p = .0277$). Significant differences were observed between the Coin ($M=1.25$) and Battle ($M=1.06$) scenarios ($Z = 23.0$, $p = .0332$).
        \item Relationship with Peers (Relation PP) varied significantly ($\chi^2(2) = 10.7579$, $p = .00461$). Noteworthy contrasts emerged among the Coin ($M=2.56$) and Station ($M=2.06$) scenarios ($Z = 48.0$, $p = .00152$), as well as between the Coin and Battle ($M=2.07$) scenarios ($Z = 51.0$, $p = .00193$). 
        \item Overall involvement in the activity (Involvement) showed significant variation across scenarios ($\chi^2(2) = 9.6923$, $p = .00786$). Particularly, a significant difference was observed between the Coin ($M=4.90$) and Battle ($M=4.98$) scenarios ($Z = 8.5$, $p = .02509$).
        \item Prompted social responses to peers (SR Peers) significantly varied across VR game scenarios ($\chi^2(2) = 13.8588$, $p = .00098$). Significant differences were observed between the Coin ($M=0.05$) and Station ($M=0.04$) scenarios ($Z = 69.5$, $p = .03085$), as well as between the Coin and Battle ($M=0.05$) scenarios ($Z = 52.0$, $p = .00289$), and between the Station and Battle scenarios ($Z = 6.0$, $p = .02248$).
        \item The mean HR showed significant differences across VR game scenarios ($\chi^2(2) = 6.1587$, $p = .04599$). Notably, a significant contrast emerged between the Station ($M=0.55$) and Battle ($M=0.39$) scenarios ($Z = 63.0$, $p = .02254$).
        \item The HR SD exhibited significant differences among VR game scenarios ($\chi^2(2) = 15.0154$, $p = .00055$). Notably, significant contrasts were observed between the Coin ($M=1.25$) and Station ($M=1.32$) scenarios ($Z = 151.0$, $p = .00424$), as well as between the Coin and Battle ($M=1.32$) scenarios ($Z = 160.0$, $p = .01871$).
        \item The CV of RR intervals displayed significant differences across VR game scenarios ($\chi^2(2) = 15.2973$, $p = .00048$). Particularly, noteworthy differences were evident between the Coin ($M=-0.85$) and Station ($M=1.82$) scenarios ($Z = 212.0$, $p = .00274$), as well as between the Coin and Battle ($M=0.80$) scenarios ($Z = 229.0$, $p = .01496$).
    \end{itemize}

\end{itemize}

\begin{figure}[ht]
    \centering
    \includegraphics[width=0.6\textwidth]{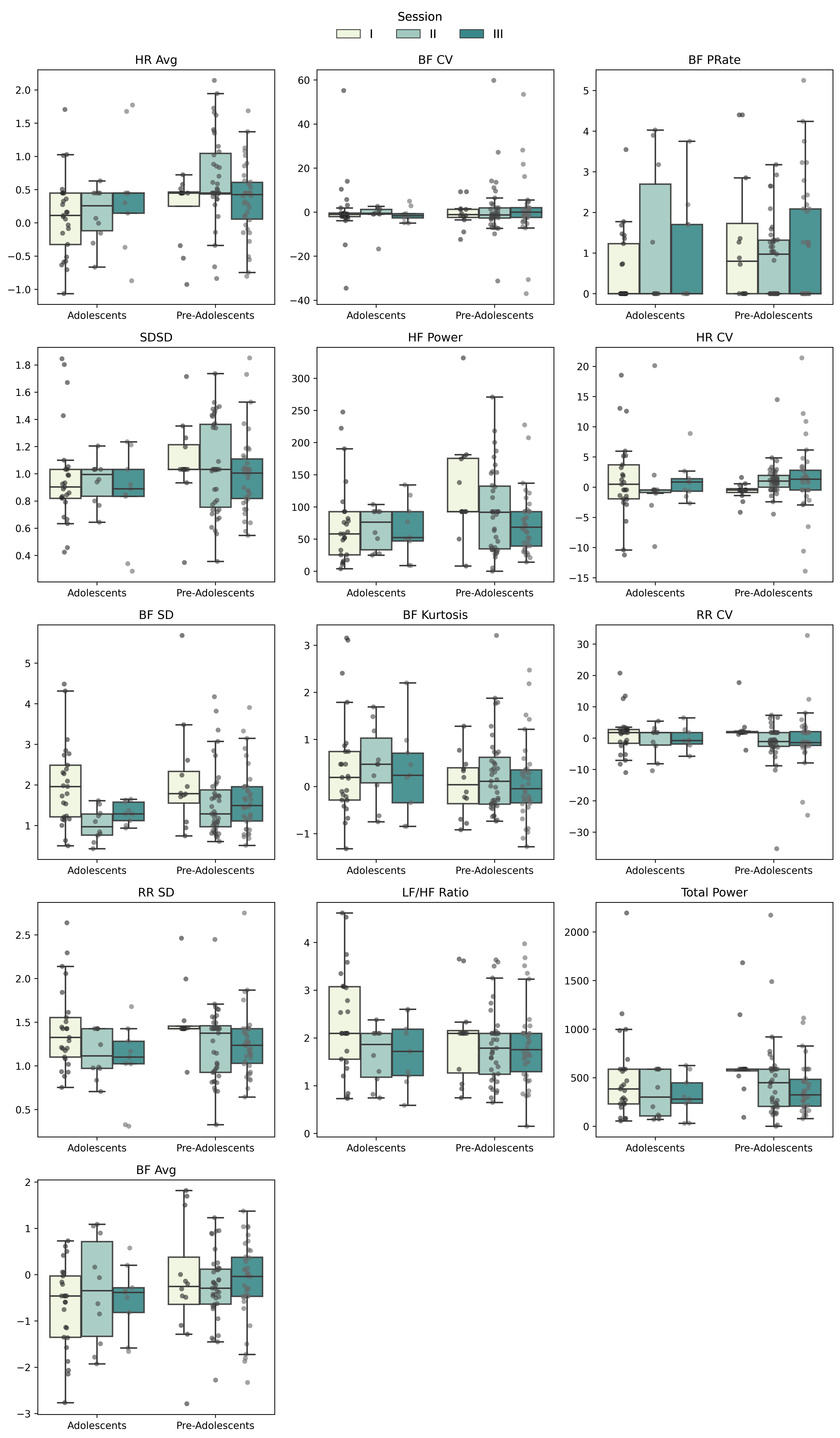} % Replace 'example-image' with the filename of your image
    \caption{Physiological features from significant comparison between sessions (I, II, III), divided by age sub-samples (Adolescents, Pre-Adolescents)}
    \label{fig:session}
\end{figure}

Next, we compared the 2 samples between sessions employing the same procedure to provide further insights into the impact of experienced variations on the observed metrics (Figure \ref{fig:session}).

\begin{itemize}
    \item \textbf{Adolescents:}
    \begin{itemize}
        \item For visuo-spatial difficulties (VS Diff), significant differences were observed among sessions ($\chi^2(2) = 10.333$, $p = 0.00570$). Posthoc Wilcoxon tests revealed a significant decrease from session 1 to session 3 ($Z = 0.0$, $p = 0.02314$, mean session 1 $= 1.05$, mean session 3 $= 1.00$), with average values indicating subtle variations in performance.
        \item For social openings to the therapist (SO Therapist), significant differences were observed among sessions ($\chi^2(2) = 14.732$, $p = 0.00063$). Noteworthy contrasts emerged between session 1 and session 3 ($Z = 74.5$, $p = 0.01783$, mean session 1 $= 2.57$, mean session 3 $= 2.78$), indicating an increase from session 1 to session 3.
        \item For overall involvement in the activity (Involvement), significant differences were observed among sessions ($\chi^2(2) = 8.73$, $p = 0.01273$). Notably, a significant increase in engagement was found from session 1 to session 3 ($Z = 0.0$, $p = 0.00975$, mean session 1 $= 4.90$, mean session 3 $= 5.00$), indicating a heightened level of involvement over the course of the sessions.
        \item Considering the rates of prompted social responses to peers (SR Peers), there were significant differences observed among sessions ($\chi^2(2) = 24.10$, $p < 0.00001$). Posthoc Wilcoxon tests revealed a notable increase in prompted social responses from session 1 to session 3 ($Z = 24.0$, $p = 0.00109$, mean session 1 $= 0.04$, mean session 3 $= 0.06$).
        \item Also, if we consider prompted social responses to the therapist (SR Therapist), significant differences were detected among sessions ($\chi^2(2) = 16.75$, $p = 0.00023$). Posthoc Wilcoxon tests revealed a notable increase in prompted social responses between session 1 and session 3 ($Z = 33.0$, $p = 0.00268$, mean session 1 $= 0.02$, mean session 3 $= 0.03$).
        \item For HR Mean, a significant difference was found among sessions ($\chi^2(2) = 9.06$, $p = 0.01077$). Posthoc Wilcoxon tests revealed a noteworthy increase in heart rate mean between session 1 and session 3 ($Z = 75.5$, $p = 0.01923$, mean session 1 $= 0.24$, mean session 3 $= 0.44$), indicating an elevation in heart rate over the course of the sessions.
        \item Considering BF CV, significant differences were observed among sessions ($\chi^2(2) = 11.23$, $p = 0.00364$). Posthoc Wilcoxon tests indicated a significant increase in breathing frequency coefficient of variation between session 1 and session 3 ($Z = 81.0$, $p = 0.00946$, mean session 1 $= 3.88$, mean session 3 $= 7.01$).
        \item Considering the BF peaks rate (BF PRate), significant differences were observed among sessions ($\chi^2(2) = 9.41$, $p = 0.00905$). Posthoc Wilcoxon tests revealed a significant increase in breathing frequency peak rate between session 1 and session 3 ($Z = 65.5$, $p = 0.01427$, mean session 1 $= 0.85$, mean session 3 $= 1.22$).
        \item For SDSD, significant differences were observed among sessions ($\chi^2(2) = 6.10$, $p = 0.04739$). Posthoc Wilcoxon tests revealed a significant increase in the SDSD of RR intervals between session 1 and session 3 ($Z = 60.0$, $p = 0.00581$, mean session 1 $= 0.94$, mean session 3 $= 1.02$).
        \item Considering HF power, significant differences were observed among sessions ($\chi^2(2) = 8.44$, $p = 0.01472$). Posthoc Wilcoxon tests revealed a significant increase in the HF power of RR intervals between session 1 and session 3 ($Z = 88.0$, $p = 0.02626$, mean session 1 $= 72.31$, mean session 3 $= 86.88$).
        \item Considering LF/HF ratio, significant differences were observed among sessions ($\chi^2(2) = 8.96$, $p = 0.01132$). Posthoc Wilcoxon tests revealed a significant decrease between session 1 and session 3 ($Z = 81.0$, $p = 0.02831$, mean session 1 $= 2.38$, mean session 3 $= 2.01$).
    \end{itemize}
    
    \item \textbf{Adolescents:}
    \begin{itemize}
        \item Considering social openings to the therapist (SO Therapist), significant differences were observed among sessions ($\chi^2(2) = 10.09$, $p = 0.00645$). Posthoc Wilcoxon tests revealed a significant decrease in social openings rates between session 1 and session 3 ($Z = 188.0$, $p = 0.00157$, mean session 1 $= 3.09$, mean session 3 $= 2.08$).
        \item Considering prompted social openings to peers (SO Peers), significant differences were observed among sessions ($\chi^2(2) = 17.30$, $p = 0.00018$). Posthoc Wilcoxon tests indicated a significant decrease in scores for social interaction between session 1 and session 3 ($Z = 115.0$, $p = 0.00072$, mean session 1 $= 0.23$, mean session 3 $= 0.14$).
        \item For prompted social responses to peers (SR Peers), significant differences were found among sessions ($\chi^2(2) = 31.04$, $p < 0.0001$). Posthoc Wilcoxon tests revealed a significant decrease in reciprocal social interaction rates from session 1 to session 3 ($Z = 97.0$, $p = 0.00006$, mean session 1 $= 0.08$, mean session 3 $= 0.04$).
        \item For prompted social responses to the therapist (SR Therapist), significant differences were observed among sessions ($\chi^2(2) = 36.36$, $p < 0.0001$). Posthoc Wilcoxon tests revealed a significant decrease in reciprocal social responses to the therapist from session 1 to session 3 ($Z = 52.5$, $p < 0.0001$, mean session 1 $= 0.03$, mean session 3 $= 0.01$).
        \item For HR CV, significant differences were found among sessions ($\chi^2(2) = 10.31$, $p = 0.00578$). Posthoc Wilcoxon tests revealed a significant increase in HR CV from session 1 to session 3 ($Z = 220.0$, $p = 0.00636$, mean session 1 $= -0.51$, mean session 3 $= 0.69$).
        \item For average BF, a significant difference was detected among sessions ($\chi^2(2) = 6.68$, $p = 0.03547$). Posthoc Wilcoxon tests indicated a significant decrease in breathing frequency from session 1 to session 3 ($Z = 279.0$, $p = 0.04962$, mean session 1 $= -0.39$, mean session 3 $= -0.19$).
        \item For SD of BF, there was a significant difference among sessions ($\chi^2(2) = 11.66$, $p = 0.00294$). Posthoc Wilcoxon tests revealed a significant decrease in breathing frequency SD from session 1 to session 3 ($Z = 279.0$, $p = 0.04962$, mean session 1 $= 2.00$, mean session 3 $= 1.80$).
        \item For BF CV, significant differences were found among sessions ($\chi^2(2) = 14.14$, $p = 0.00085$). Posthoc Wilcoxon tests revealed a significant decrease between session 1 (M $= 6.94$) and session 3 (M $= 3.45$) ($Z = 218.0$, $p = 0.00589$).
        \item For RR intervals CV, a significant difference was found among sessions ($\chi^2(2) = 18.68$, $p = 0.00009$). Posthoc Wilcoxon tests revealed a significant decrease between session 1 (M $= 2.01$) and session 3 (M $= 0.36$) ($Z = 220.0$, $p = 0.00637$).
        \item For RR intervals SD, there was a significant difference among sessions ($\chi^2(2) = 6.92$, $p = 0.03147$). Posthoc Wilcoxon tests revealed a significant decrease between session 1 (M $= 1.45$) and session 3 (M $= 1.30$) ($Z = 154.0$, $p = 0.00034$).
        \item For Total Power, there was a significant difference among sessions ($\chi^2(2) = 6.92$, $p = 0.03147$). Posthoc Wilcoxon tests revealed a significant decrease between session 1 (M $= 602.06$) and session 3 (M $= 447.45$) ($Z = 142.0$, $p = 0.00018$).
        \item For HF power, significant differences were found among sessions ($\chi^2(2) = 12.19$, $p = 0.00226$). Posthoc Wilcoxon tests revealed a significant decrease in HF power between session 1 (M $= 100.34$) and session 3 (M $= 79.90$) ($Z = 188.0$, $p = 0.00167$).
\end{itemize}

\end{itemize}

\subsection{Relationship Between Behavioral and Physiological Variables}

\subsubsection{Physiological-Behavioral Correlations}
The GLM analysis revealed several noteworthy findings regarding the relationship between behavioral and physiological variables. Before running the models, we filtered the physiological features by computing the Variance Inflation Factor (VIF) to check for multicollinearity. Features with VIF greater than 5 were excluded, which resulted in excluding RR SD, SDSD, and RMSSD.

\begin{itemize}
    \item For spontaneous social openings to peers (SO Peers), analyzed with a Gamma GLM ($n=99$), BF Kurtosis emerged as a significant predictor (coef $= -0.1353$, $p = .033$). The model demonstrated good quality of fit, indicated by a deviance $= 50.249$, an AIC $= 595.205$, and a Pseudo R-squared $= 0.1830$, reflecting a moderate explanation of variance within the data.
    \item For spontaneous social responses to peers (SR Peers), analyzed with a Gamma GLM ($n=91$), the model identified BF Kurtosis (coef $= -0.1876$, $p = .003$) and RR CV (coef $= -0.0125$, $p = .011$) as significant predictors. The model demonstrated a good fit, with a deviance $= 39.505$, an AIC $= 445.47$, and Pseudo R-squared $= 0.2128$, indicating the model's adequacy in capturing the data's structure.
    \item In analyzing social openings to the therapist (SO Therapist) with a Gamma GLM ($n=85$), the model highlighted BF Kurtosis (coef $= -0.2393$, $p < .001$) as a significant predictor, indicating its substantial influence. The fit was quantified by a deviance $= 40.211$, an AIC $= 388.96$, and  Pseudo R-squared $= 0.284$ suggesting adequacy in capturing the response variable's variability.
    \item In analyzing spontaneous social responses to the therapist (SR Therapist) with a Gamma GLM ($n=96$), LF/HF Ratio emerged as a significant factor (coef $= 0.1630$, $p = .019$), demonstrating its influence. The model showed good fit, evidenced by a deviance $= 29.55$, an AIC $= 417.16$, and a Pseudo R-squared $= 0.2162$, indicating a reasonable explanation of variance.
    \item Analyzing Involvement level with a Gamma GLM ($n=127$) observations revealed a highly precise model. Key predictors included HR SD (coef $= -0.0800$, $p = .014$) and BF CV (coef $= -0.0002$, $p = .034$), with LF/HF ratio also showing significance (coef $= -0.0198$, $p = .046$). The model showed a deviance $= 1.7832$, an AIC $= 247.54$, and it achieved a Pseudo R-squared $= 0.1185$, denoting a modest explanatory capacity.
    \item Analyzing visuo-spatial difficulties (VS Diff) with a Gamma GLM ($n=131$) yielded a precise model fit, evidenced by a deviance $= 23.026$, an AIC $=  254.43$, and Pseudo R-squared $= 0.1733$, showing a modest explanatory capacity. The model's significance was notably highlighted by RR mean (coef $= -0.2234$, $p = .048$).
    \item VR adaptation difficulties (Adaptation Diff) was analyzed using a Gamma GLM ($n=129$), showing a strong model fit with a deviance $= 18.43$, an AIC $= 216.575$, and a Pseudo R-squared $= 0.2192$, denoting a modest explanatory capacity. The model featured significant predictors such as HR CV (coef $= -0.0076$, $p = .016$), RR Kurtosis (coef $= -0.0232$, $p = .039$), and Total Power (coef $= 0.0003$, $p = .005$), indicating their roles in adaptation to virtual reality environments.
    \item The quality of relationship between peers (Relation PP) was analyzed with a Gamma GLM ($n=133$), resulting in a robust fit, as shown by a deviance $= 32.413$, an AIC $= 450.156$, and a Pseudo R-squared $= 0.1352$, denoting a modest explanatory capacity. The model pinpointed BF SD (coef $= -0.0699$, $p = .004$) and RR CV (coef $= -0.0033$, $p = .026$) as significant predictors, indicating their influence on relationship behaviors.
\end{itemize}

\subsubsection{Unsupervised Analysis of Behavioral and Physiological Relationships}

\begin{figure}[ht]
    \centering
    \includegraphics[width=1\textwidth]{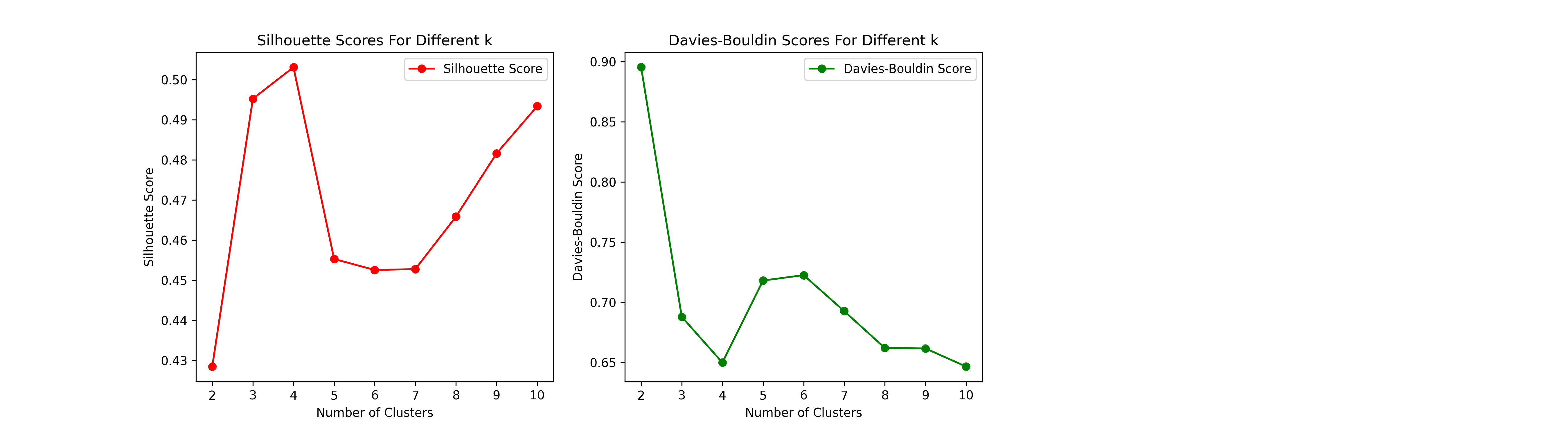} % Replace 'example-image' with the filename of your image
    \caption{Silhouette and Davies-Bouldin Scores for different K}
    \label{fig:silhouette}
\end{figure}

The unsupervised clustering analysis yielded significant insights into the interplay between behavioral and physiological variables. From variance analysis we reduced the total features (N=20) to a core set of 10 features: visuo-spatial difficulties, spontaneous social openings to peers, spontaneous social responses to peers, social openings to therapist, spontaneous social responses to therapist’, VR adaptation difficulties, quality of peers’ relationship, involvement, age, and subject ID. This reduction was applied to isolating the most informative features for subsequent analyses. Advancing to dimensionality reduction, we employed t-SNE to extract 2 components and we applied KMeans on the embedding. For optimizing cluster analysis, our results revealed that a configuration of 3 clusters represents the optimal balance, maximizing the Silhouette score while simultaneously minimizing the Davies-Bouldin score (Figure \ref{fig:silhouette}). This configuration ensures an effective delineation of the sample data into distinct groups, without resulting in overly fragmented clusters.\\

\begin{figure}[ht]
    \centering
    \includegraphics[width=0.7\textwidth]{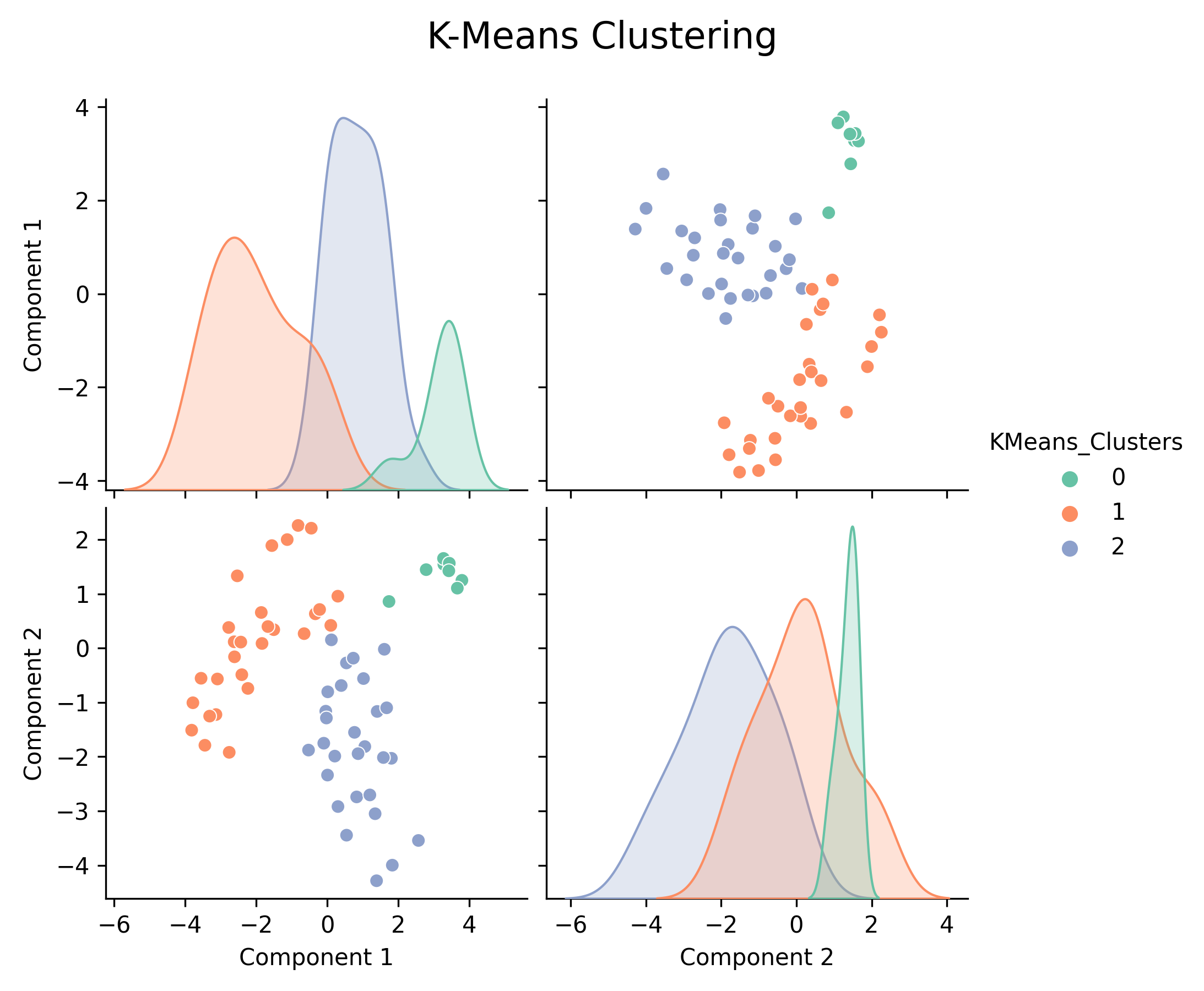} % Replace 'example-image' with the filename of your image
    \caption{KMeans Clustering results on t-SNE embedding (2 components)}
    \label{fig:clustering}
\end{figure}

Upon applying the K-Means algorithm, we successfully delineated 3 distinct clusters: Cluster 0 (N=8), Cluster 1 (N=28), and Cluster 2 (N=28) (Figure \ref{fig:clustering}). This configuration yielded a Silhouette score = 0.495 and a Davies-Bouldin Index = 0.687, coupled with an elbow coefficient = 133.77, collectively indicating the presence of well-defined and suitably separated clusters.\\

\begin{figure}[ht]
    \centering
    \includegraphics[width=0.8\textwidth]{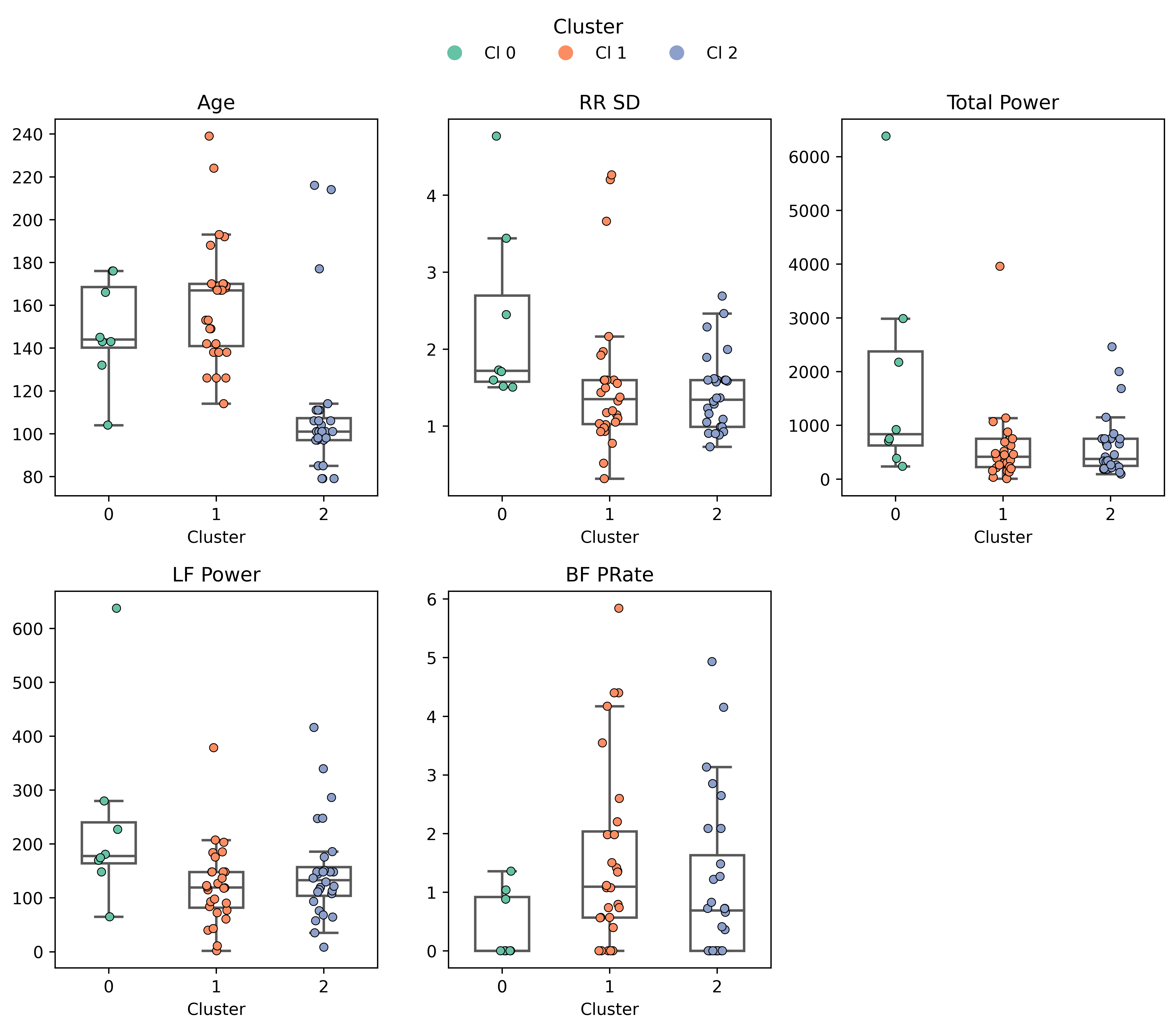} % Replace 'example-image' with the filename of your image
    \caption{Age and physiological clusters profile from significant comparisons}
    \label{fig:clusters}
\end{figure}

Finally, we validated the three clusters by analyzing the characteristics of the emerging subgroups with paired comparisons between cluster pairs (Table \ref{tab:clusters_characteristics}, Figure \ref{fig:clusters}). The analysis revealed significant differences when comparing Cluster 0 with Cluster 1 in several dimensions:

\begin{itemize}
    \item Spontaneous social openings ($U = 206.0$, $p < .001$) and spontaneous social responses to peers ($U = 168.0$, $p = .034$) were higher in Cluster 1 compared to Cluster 0.
    \item Prompted social responses to the therapist ($U = 84.0$, $p = .0084$) were higher in Cluster 0.
    \item VR adaptation difficulties ($U = 2.5$, $p < .001$) and visuo-spatial difficulties ($U = 14.5$, $p < .001$) were higher in Cluster 0, together with lower attention to instruction ($U = 167$, $p = .0336$).
    \item Involvement ($U = 224.0$, $p < .001$) and relationship between peers ($U = 200.0$, $p < .001$) were significantly higher in Cluster 1.
    \item Considering physiological measures, Cluster 0 exhibited a greater autonomic response, indicated by higher Total Power ($U = 53.0$, $p = .0258$) and Low-Frequency power ($U = 50.0$, $p = .0191$), alongside increased RR SD ($U = 48.0$, $p = .0155$), pointing to heightened autonomic activity in Cluster 0. Conversely, breathing Peak Rates ($U = 171.0$, $p = .024$) were more pronounced in Cluster 1.
\end{itemize}

Considering Cluster 2 in comparison to Cluster 0, the analysis revealed significant differences in:

\begin{itemize}
    \item Age showed a significant difference ($U = 31.5$, $p = .0022$), with Cluster 2 presenting a younger demographic (M = 109.32) compared to Cluster 0 (M = 148.13).
    \item Visual-Spatial Difficulties ($U = 22.5$, $p < .001$) and VR Adaptation difficulties ($U = 7.5$, $p < .001$) were more pronounced in Cluster 0. 
    \item Spontaneous Social Openings with Peers ($U = 221.0$, $p < .001$), Engagement with Activities ($U = 224.0$, $p < .001$), and Quality of Relationship ($U = 190.0$, $p = .0015$) were significantly higher in Cluster 2.
    \item Considering physiological measures, Cluster 2 exhibited a less pronounced autonomic response compared to Cluster 0, as indicated by lower RR SD ($U = 44.0$, $p = .0102$) and Total Power ($U = 56.0$, $p = .0345$). This suggests a more regulated autonomic activity in Cluster 2, in contrast to the heightened responses observed in Cluster 0.
\end{itemize}

Finally by comparing Cluster 2 with Cluster 1, the analysis revealed significant differences in:
\begin{itemize}
    \item A significant age difference was observed, with Cluster 2 being younger (M = 109.32) compared to Cluster 1 (M = 161.29; $U = 75.5$, $p < .001$).
    \item Cluster 2 showed higher Spontaneous Social Openings with Peers ($U = 625.0$, $p < .001$) and Spontaneous Social Openings with Therapist ($U = 570.0$, $p = .0036$).
    \item There was a higher level of Spontaneous Social Responses to Therapist ($U = 637.0$, $p < .001$) and Prompted Social Responses to Therapist ($U = 448.0$, $p = .0417$) responses to therapists in Cluster 2.
    \item Quality of the Relation ($U = 180.5$, $p = .0004$) appeared higher in Cluster 1.
    \item Cluster 1 demonstrated fewer Difficulties in Visuo-Spatial processing ($U = 562.5$, $p < .001$) and VR Adaptation ($U = 545.5$, $p = .0029$) than Cluster 2.
    \item Considering physiological profile, no significant differences emerged.
\end{itemize}

\begin{table}[ht]
\centering
\caption{Clusters’ characteristics from significant comparisons}
\label{tab:clusters_characteristics}
\begin{tabular}{lccc}
\toprule
& \textbf{Cluster 0} & \textbf{Cluster 1} & \textbf{Cluster 2} \\
\midrule
\textbf{Feature} & & & \\
\midrule
\textbf{N} & 8 & 28 & 28 \\
\textbf{Age (months), mean (SD)} & 148.1 (24) & 161.3 (29) & 109.3 (35) \\
\textbf{Spontaneous SO Peers, mean (SD)} & 0.30 (0.8) & 4.27 (3.4) & 10.69 (8.8) \\
\textbf{Spontaneous SR Peers, mean (SD)} & 1.56 (2.4) & 4.62 (4.1) & 2.64 (1.8) \\
\textbf{SO Therapist, mean (SD)} & 2.39 (2.7) & 1.80 (2.1) & 4.72 (5.0) \\
\textbf{Spontaneous SR Therapist, mean (SD)} & 2.81 (2.4) & 1.99 (1.6) & 4.66 (3.5) \\
\textbf{Prompted SR Therapist, mean (SD)} & 0.11 (0.3) & 0.00 (0.0) & 0.08 (0.2) \\
\textbf{Adaptation Diff, mean (SD)} & 4.0 (1.1) & 1.2 (0.5) & 1.5 (0.7) \\
\textbf{VS Diff, mean (SD)} & 3.6 (1.2) & 1.0 (0.2) & 1.6 (0.8) \\
\textbf{Instructions, mean (SD)} & 3.5 (1.2) & 4.5 (0.5) & 4.2 (0.8) \\
\textbf{Involvement, mean (SD)} & 2.3 (0.5) & 4.8 (0.4) & 4.8 (0.4) \\
\textbf{Relation PP, mean (SD)} & 1.4 (0.5) & 3.2 (1.2) & 2.2 (0.5) \\
\textbf{RR SD, mean (SD)} & 2.34 (1.3) & 1.57 (1.0) & 1.41 (0.5) \\
\textbf{TP, mean (SD)} & 1816.9 (2072) & 578.2 (727) & 614.5 (579) \\
\textbf{LF, mean (SD)} & 235.2 (174) & 123.3 (73) & 147.9 (90) \\
\textbf{PRate, mean (SD)} & 0.41 (0.6) & 1.58 (1.6) & 1.08 (1.4) \\
\bottomrule
\end{tabular}
\end{table}

\section{Discussion}

This study conducted an exploratory analysis to investigate the relationship between behavioral and physiological responses during VR sessions among adolescents with ASD. This investigation addresses a gap in existing research regarding the use of physiological and behavioral metrics in VR interventions for ASD. Given the communication challenges characteristic of ASD, where individuals may not readily express their internal experiences, our analysis sought to leverage VR to provide novel insights into non-verbal, physiological manifestations of social and emotional responses. The goal was to explore the feasibility of this approach for uncovering latent behavioral cues, thereby contributing to a more comprehensive understanding of ASD behaviors within interactive VR environments.\\
The participants played three virtual game scenarios designed to train specific skills, especially: Coin Search scenario, characterized by a need for collaboration and a focus on searching for coins, aimed  to foster cooperative behavior, sustained attention and social coordination; Station scenario, which involves talking to a therapist and role playing in a structured environment as well as working memory; Battle scenario, described as the more enjoyable one where participants engage in dynamic and collaborative actions.\\
Considering preliminary control analyses, the significant sex differences observed in listening to instructions, social responsiveness, and clinical profiles such as the Visual Spatial Index (VSI) and Working Memory Index (WMI) underscore the importance of considering sex as a critical variable in the assessment and intervention planning for individuals with ASD. Specifically, the findings that males scored higher in listening to instructions while females showed greater social responsiveness and higher scores in PRI and WMI highlight the diverse needs and strengths of males and females with ASD. These differences may reflect underlying neurodevelopmental and psychological mechanisms that influence cognitive processing, social cognition, and interaction styles \cite{di2024characterizing}. These results align with existing literature suggesting sex-specific phenotypic expressions of ASD \cite{mcdonnell2021sex}. The negative correlations between age and physiological measures such as heart rate, heart rate variability (HRV) indices, and high-frequency (HF) power, coupled with the positive correlation with the low-frequency/high-frequency (LF/HF) ratio, suggest age-related changes in autonomic nervous system function. Considering behavioral profile, the observed shifts in social behavior, with older participants showing fewer spontaneous social openings but better quality of relationships with peers, may indicate a developmental trajectory in which adolescents with ASD become more selective in their social interactions, possibly as a result of accumulating social experiences or maturation of social cognition. \\
Given the significant correlations between age and various behavioral and physiological measures, our dataset was split into two age groups (pre-adolescents, adolescents). This stratification was informed by the observed correlations, allowing for a more nuanced investigation into how developmental stages may impact the efficacy of VR interventions and the expression of ASD characteristics.

\subsection{Scenarios}

In exploring the dynamics of different VR gaming environments and activities on adolescents and pre-adolescents with ASD, we assessed behavioral and physiological responses across different VR scenarios: Coin, Station, and Battle. \\
Among adolescents, significant differences were observed in physiological responses (e.g., BF SD and RR CV) across VR scenarios, suggesting that different VR environments elicit varying levels of physiological arousal. The observed increase in HRV from the Coin scenario to the Station scenario, as indicated by the significant changes in the RR CV, may suggest a state of reduced stress or heightened parasympathetic activity, which is associated with relaxation and recovery \cite{cheng2020heart,shaffer2017overview}. This finding is consistent with the design of the game scenario and the socio-cognitive tasks provided. In fact, the Station scenario requires role-playing and dyadic interaction with a therapist in a structured way; whereas, the Coin scenario requires more spontaneous interactions with all group members with higher peer-to-peer collaboration dynamics. This proactive engagement reflects the scenario's success in encouraging communication within a supportive context, providing a valuable opportunity to practice and enhance social skills. The decrease in the quality of peer relationships from the Coin to Station and Battle scenarios confirm previous results. The shift towards tasks that require individual or dyadic focus, as in the Station and Battle scenarios, may temporarily affect adolescents' ability to engage positively with peers. However, this observation also emphasizes the need for VR interventions to carefully design scenarios that not only promote individual skill development but also foster positive peer interactions.\\
Considering pre-adolescents, the physiological data reveal variations across the scenarios that potentially hint at differing levels of arousal. The notable increase in both the CV of RR intervals, and the SD of HR moving from the Coin to Station and Battle scenarios mirrors the trends seen in adolescents, suggesting an increase in HRV. Unlike adolescents, the VR adaptation difficulties showcased variability, with pre-adolescents finding the Battle scenario less challenging to adapt to compared to Coin. This ease of adaptation might be attributable to the inherently engaging nature of the Battle scenario, which, being action-oriented, could captivate the participants' attention and interest, facilitating smoother engagement without the heightened stress potentially induced by more complex, collaborative tasks seen in Coin. Similarly, the significant variation in overall Involvement in the activity across scenarios, particularly the significant difference between the Coin and Battle scenarios, reinforces the idea that engagement levels can be deeply influenced by the thematic and interactive elements of the VR environment. The temporal sequence of the Battle scenario at the end of the session could also bias these differences. As participants progress through the VR session, starting with Coin and moving through Station to Battle, they gradually become more accustomed to the VR environment and more involved in the activities, not just because of its engaging nature but also due to increased comfort and familiarity with the VR setting.

\subsection{Sessions}
In analyzing the progression and differences of VR sessions we compared physiological and behavioral measures over the course of the 3 sessions. The physiological data from adolescents sub-sample, encompassing heart rate (HR Mean), breathing frequency variability (BF CV), peak breathing rates (BF PRate), SD of successive differences (SDSD), HF power, and the LF/HF ratio, collectively indicate an increase in physiological arousal and HRV from the first to the third session. Specifically, the notable increase in HR Mean and the significant changes in HRV metrics such as SDSD, HF power, and a decrease in the LF/HF ratio seem to underscore a nuanced adaptation process. These physiological markers, particularly the increase in HRV (as evidenced by HF power and a reduced LF/HF ratio), could suggest a state of heightened engagement coupled with reduced stress over time \cite{di2017monitoring,shaffer2017overview}. This trend would be consistent with the understanding that HRV can be a marker of the body's ability to adapt and manage stress, indicating that adolescents become more physiologically attuned and potentially more relaxed in the VR environment with repeated exposure. Behaviorally, adolescents showed significant improvements in visuo-spatial difficulties, social openings (both to therapists and peers), and overall involvement in the activity, with notable advancements from the initial to the final session. The increase in social openings to therapists and peers, along with better instruction listening and higher involvement levels, suggests that adolescents become more comfortable, engaged, and socially interactive as they acclimate to the VR setting. This progression seems to imply that structured and repeated VR sessions can foster an environment conducive to social skill development and increased participation, which are critical components of interventions for ASD. \\
Considering pre-adolescents, the increase in HRV suggests enhanced physiological adaptation and potentially reduced stress levels, a pattern that aligns with findings in adolescents, indicating a consistent physiological response to VR exposure across age groups. This cross-age consistency in HRV enhancement supports the notion that VR interventions can foster an environment conducive to stress modulation and physiological regulation, irrespective of the developmental stage. However, the decrease in high-frequency (HF) power observed among pre-adolescents requires careful consideration. Unlike the general trend of increased physiological regulation, this decrease could suggest a more complex or varied response to VR exposure in younger participants compared to adolescents. Considering behavioral profile, the overall decrease in prompted social interaction rates among pre-adolescents, requires a deeper consideration of the implications of these changes. The decline in prompted social interactions over time does not necessarily indicate a reduction in overall social engagement or ability. Instead, it could suggest a developmental progression: as pre-adolescents become more accustomed to the VR environment and its social dynamics, they may rely less on external prompts for interaction, reflecting an increase in spontaneous social engagement and a growing sense of independence. This shift towards spontaneous social interactions, without the need for prompts, suggests an improvement in natural social initiative and comfort within the VR setting. Similarly, the observed decrease in social openings to the therapist could indicate a reduced need for support, as pre-adolescents gain confidence and competence in navigating the VR environment independently.These findings underscore the importance of considering developmental stages when designing and implementing VR interventions for ASD. The capacity of VR to support physiological regulation and promote adaptive social behaviors across different age groups highlights its value as a versatile and impactful therapeutic tool. Moving forward, tailoring VR content to encourage not only engagement and stress management but also autonomous social interaction will be key in maximizing the benefits of VR for individuals with ASD.

\subsection{Intra-Session}

This session-level analysis transcends age distinctions, aiming to elucidate the broader impacts of VR therapy on physiological and behavioral dimensions across all sessions and participants. The HRV-related metrics emerged as interesting in exploring the dynamics of social interaction and adaptation within VR settings. Notably, the inverse correlation between RR intervals CV and spontaneous social responses and quality of relation with peers suggests that higher levels of physiological arousal or variability—often signs of discomfort—may deter social engagement. Specifically, the decrease in HRV when increasing social-related behaviors underscores the sensitivity of individuals with ASD to the social demands of VR environments. These findings align with the broader understanding of ASD, where social situations can escalate stress, thereby modulating social responsiveness through autonomic regulation. Beyond social behaviors, HRV metrics suggest an influence on broader aspects of VR engagement, including involvement levels as well as adaptation difficulties within VR settings. For instance, HR SD and HR CV seem inversely correlated with a higher level of Involvement in the activities and more adaptation difficulties respectively. Increased variability in heart rate, may suggest fluctuations in attention or emotional arousal, contributing to greater effort to involve in the activity and challenges in adapting to the demands of VR environments.\\
The LF/HF ratio and Total Power offer insights into autonomic balance and overall HRV amplitude, respectively, revealing their implications for behavioral features. The LF/HF ratio, representing sympathovagal balance \cite{shaffer2017overview}, may indicate the individual's physiological stress response or arousal level during social interactions and VR experiences. Consistently with results from HRV features, a higher LF/HF ratio was found to be associated with higher rates of social responses to the therapist, confirming the physiological effort or discomfort involved in having social interactions for autistic individuals. However, we found negative correlations observed between both HRV metrics and LF/HF Ratio with the Involvement levels in activities. This discrepancy between HRV metrics and LF/HF Ratio may indicate a more complex relationship between autonomic regulation and engagement. The positive correlation between Total Power and adaptation difficulties within VR environments suggests that individuals with higher overall HRV amplitude may experience greater challenges in adapting \cite{shaffer2017overview}. In the context of VR settings, higher Total Power may suggest more pronounced physiological responses to virtual stimuli, potentially leading to difficulties in adapting to virtual environments, consistent with HRV dynamics.\\
Finally, the salience of Breathing Frequency variability (BF Kurtosis, BF Coefficient Variation) across multiple social behavior measures underscores the intricate interplay between respiratory dynamics and socioemotional processes in individuals with ASD. Variations in breathing patterns can serve as valuable indicators of underlying emotional states, cognitive processes, and physiological arousal levels. For autistic individuals, who often experience challenges in regulating emotions and responding to social cues, fluctuations in breathing frequency may reflect heightened physiological arousal or stress in social contexts. These variations in respiratory dynamics could influence the individual's ability to interpret and respond to social cues, thereby shaping their interactions with others and impacting their overall socioemotional well-being. The observed associations between BF variability and social behaviors highlight the importance of considering respiratory dynamics as integral components of the autonomic regulation framework in ASD populations. By examining breathing frequency variability alongside traditional HRV metrics, researchers and clinicians can gain a more comprehensive understanding of the physiological underpinnings of social behaviors in individuals with ASD. This holistic approach to assessing autonomic function may facilitate the development of targeted interventions aimed at enhancing emotional regulation, social communication, and interpersonal skills in individuals with ASD.

\subsection{Clustering}
The integration of unsupervised clustering analysis with paired comparisons between clusters offers valuable insights into the complex relationship between behavioral profiles, physiological features, and demographic characteristics among individuals with ASD. This approach allows us to identify distinct subgroups within the sample population and explore how these subgroups differ in terms of both behavioral manifestations and physiological responses. Our exploratory analysis offers a preliminary yet insightful glimpse into the complex interplay between behavioral and physiological responses to VR-based interventions among autistic individuals.\\
We found 3 well-defined session clusters, each telling a unique story about the interplay between VR environments and ASD. Cluster 0 \textit{Challenging-Sessions}, although smaller in size, represents a subgroup characterized by unique behavioral and physiological profiles. Compared to Clusters 1 and 2, Cluster 0 exhibits higher levels of VR adaptation difficulties, visual-spatial processing challenges, and prompted social responses to the therapist, together with lower rates of spontaneous social openings to peers. Additionally, Cluster 0 demonstrates heightened autonomic activity, as evidenced by increased Total Power, Low-Frequency power, and RR SD. Physiologically, the heightened autonomic response observed suggests these challenge-intensive sessions may induce stress or heightened engagement, underscoring the necessity for careful consideration of session intensity and support levels.\\
In contrast, Clusters 1 and 2, while varied in age, share a similar profile in terms of physiological responses within VR settings. \textit{Young Enriched-Sessions} grouped into Cluster 2 exhibit a heightened level of social interaction, as evidenced by higher rates of spontaneous social openings and responses. These results indicate that Cluster 2 sessions are particularly effective in engaging younger participants in social exchanges, suggesting these sessions may employ VR scenarios that encourage or necessitate social initiation and responsiveness more so than those in Cluster 1. Conversely, Cluster 1 \textit{Interaction Expert-Session} sessions demonstrate a stronger capability in minimizing VR visuo-spatial and adaptation difficulties, compared to Cluster 2. This suggests that older participants may have a better ability to navigate and adapt to VR environments, reflecting either a developmental advantage in spatial processing or the session designs' suitability to their cognitive capacities. Despite the more pronounced social initiation seen in Cluster 2, Cluster 1 sessions are marked by a higher quality of relation with peers, suggesting that while older participants may engage less frequently in spontaneous social interactions, the depth and quality of their social engagements within VR are more substantial. This could imply that sessions designed for older participants foster environments that support more meaningful social connections, possibly through scenarios that require collaborative problem-solving or sustained interaction. Notably, no significant physiological differences were observed between the two clusters, highlighting a complex landscape where behavioral engagement and social interaction enhancements do not straightforwardly correlate with physiological changes. This lack of physiological variance across age groups and social interaction skills levels points to the nuanced effects of VR sessions on participants with ASD, suggesting that the autonomic nervous system's response to VR therapy may be influenced by factors beyond immediate social and cognitive demands.

\subsection{Limitations}

A primary limitation of our research is the relatively small sample size, which constrains the breadth and depth of our analyses. Future studies would benefit from a larger, more representative sample that encompasses a broader range of demographic variables, including gender, age, and severity of ASD symptoms. Furthermore, interpreting physiological data, such as HRV and respiration patterns, within the context of VR sessions presents significant challenges, particularly due to the under-investigation of these aspects in the existing literature. Understanding the implications of these physiological responses in the context of VR-based interventions for autistic individuals remains an area requiring further exploration and research. Larger samples would not only enhance the statistical power of such studies but also allow for the investigation of more complex models of interaction between participant characteristics and VR intervention outcomes. Finally, while our session-based approach provides valuable insights into group-level responses, it may not fully capture individual variations within those sessions. This aggregation might mask nuanced behaviors or physiological responses unique to individual participants, potentially oversimplifying the complex interactions between VR environments and ASD characteristics.

\subsection{Conclusions \& Perspectives}

Our findings highlight the importance of considering both the developmental stage and the unique characteristics of ASD in designing VR interventions. The similarities and differences observed between adolescents and pre-adolescents in their responses to VR scenarios highlight the nuanced impact of virtual environments on individuals with ASD. For both age groups, scenarios that encourage communication and offer a balance between challenge and support appear to foster beneficial physiological and social responses. However, the variance in adaptation difficulties and the impact on peer relationships across scenarios and sessions indicate that VR interventions must be carefully tailored to support positive social interactions and ensure participants are neither overwhelmed nor disengaged. This balance is crucial for harnessing the therapeutic potential of VR for individuals with ASD, emphasizing the need for interventions that are adaptable to the diverse needs and responses of this population. Furthermore, by relying on physiological markers, therapists and intervention designers can gain valuable insights into the stressors and challenges faced by participants, enabling the customization of VR experiences to better support individual needs and promote social engagement. These insights collectively affirm the multifaceted impact of physiological regulation on the VR intervention experience for individuals with ASD. By shedding light on the significant role of HRV and breathing frequency variability, our findings advocate for the integration of physiological monitoring and targeted interventions within VR therapies. These physiological indicators reveal the internal dynamics related to stress and emotional regulation that might not always be observable through external behaviors. By integrating real-time physiological monitoring into VR therapies, therapists and intervention designers can gain immediate feedback on how individuals with ASD are physiologically responding to social stimuli within the virtual environment. This feedback mechanism can inform therapists about the underlying dynamics of social stimulation and engagement levels, enabling them to adapt the intervention in real-time to better suit the individual's needs.\\
Incorporating physiological features as feedback to therapists opens a new avenue for enhancing digital interventions for ASD. It allows for the identification and interpretation of subtle, yet significant, undercurrents of emotional and stress responses that are not always perceptible through visible behaviors alone. This approach enables a more dynamic and responsive intervention strategy, where adjustments can be made on-the-fly to optimize therapeutic outcomes. Such adaptive interventions, informed by physiological feedback, hold the promise of delivering more personalized, effective, and supportive therapies, tailored to the unique physiological profiles and needs of individuals with ASD.
Such approaches could profoundly enhance the efficacy of VR as a therapeutic tool, paving the way for more personalized, effective, and supportive interventions tailored to the unique needs and physiological profiles of individuals with ASD.

%Bibliography
\bibliographystyle{unsrt}  
\bibliography{references}

\end{document}